\documentclass[useAMS,usenatbib,usegraphicx,fleqn]{mn2e}			
\usepackage{times}
\usepackage{natbib} 
\usepackage[dvips]{graphicx,color}
\usepackage{amsmath}
\usepackage{amssymb}
\usepackage{textcomp}
\usepackage[normalem]{ulem}

\title[The origin of S0s in clusters]
{The origin of S0s in clusters: evidence from the bulge and disc star formation histories}
\author[E.J. Johnston, A. Arag\'on-Salamanca and M.R. Merrif\mbox{}ield]{Evelyn~J.~Johnston,$^1$\thanks{Email: Evelyn.Johnston@nottingham.ac.uk} Alfonso~Arag\'on-Salamanca$^1$ and Michael~R.~Merrif\mbox{}ield$^1$\\
  $^1$School of Physics and Astronomy, University of Nottingham, University Park, Nottingham, NG7 2RD, UK\\
}

\begin{document}

\maketitle

\begin{abstract}
  The individual star formation histories of bulges and discs of
  lenticular (S0) galaxies can provide information on the processes
  involved in the quenching of their star formation and subsequent
  transformation from spirals. In order to study this transformation
  in dense environments, we have decomposed long-slit spectroscopic
  observations of a sample of 21 S0s from the Virgo Cluster to produce
  one-dimensional spectra representing purely the bulge and disc light
  for each galaxy. Analysis of the Lick indices within these spectra
  reveals that the bulges contain consistently younger and more
  metal-rich stellar populations than their surrounding discs,
  implying that the final episode of star formation within S0s occurs
  in their central regions.  Analysis of the $\alpha$-element
  abundances in these components further presents a picture in which
  the final episode of star formation in the bulge is fueled using gas
  that has previously been chemically enriched in the disc, indicating
  the sequence of events in the transformation of these galaxies.
  Systems in which star formation in the disk was spread over a longer
  period contain bulges in which the final episode of star formation
  occurred more recently, as one might expect for an approximately
  coeval population in which the transformation from spiral to S0
  occurred at different times.  With data of this quality and the new
  analysis method deployed here, we can begin to describe this process
  in a quantitative manner for the first time.
\end{abstract}

\begin{keywords}
  galaxies: elliptical and lenticular -- 
  galaxies: evolution --
  galaxies: formation -- 
  galaxies: stellar content
\end{keywords}

\section{Introduction}\label{sec:introduction}
Lenticular galaxies (S0s) lie between spirals and ellipticals on the
Hubble Sequence, sharing the discy morphology of the spiral galaxies
but containing the redder colours and old stellar populations
generally seen in ellipticals. As a result, they are often seen as a
transitional phase between spirals and ellipticals, and so
understanding their formation is key to understanding the evolution of
galaxies and the significance of the Hubble Sequence.

Plenty of evidence exists to suggest an evolutionary link between
spirals and S0s.  The morphology--density relation of
\citet{Dressler_1980} showed that spirals tend to dominate lower
density regions of space, such as the field, while S0s begin to
dominate as you move into groups and clusters. This finding suggests
that the local environment plays a key role in the transformation of
spirals to S0s, where the higher frequency of interactions in groups
and clusters contribute significantly to the quenching of star
formation in the progenitor spirals.  Studies such as
\citet{Dressler_1997}, \citet{Fasano_2000} and \citet{Desai_2007} have
also found a link with redshift, where the fraction of S0s increased
toward lower redshifts while that of spirals decreased, thus
revealing that S0s are more common at more recent epochs than in the
past.


Many processes have been suggested to explain this transformation,
most of which focus on an interaction that quenches the star formation
in the disc followed by passive evolution as the galaxy fades to an
S0. The interaction that triggers the transformation could be with the
intra-cluster medium as the galaxy travels through the cluster, where
the disc gas is removed by ram pressure stripping \citep{Gunn_1972} or
the halo gas is stripped and the star formation quenched by starvation
\citep{Larson_1980,Bekki_2002}.  Alternatively, interactions with
neighbouring galaxies could lead to the gas being stripped by
harassment \citep{Moore_1996,Moore_1998,Moore_1999}, or minor mergers
could initiate starbursts that use up all of the remaining gas
throughout the entire disc \citep{Mihos_1994}. It is still uncertain
whether any one of these processes dominate the transformation, or if
the process changes with time, environment or luminosity. Evidence
that such variations do occur was found by
\citet{Barway_2007,Barway_2009}, who found that the transformation
process is dependent on the luminosity of the galaxy, where fainter
S0s evolved through secular processes while brighter galaxies were
created by more turbulent ones.

The different processes that have been suggested to explain the
transformation of spirals to S0s would affect the bulges and discs in
different ways, making their individual star formation histories key
to understanding the transformation process.  For this reason, many
studies have tried looking at variations in the stellar populations of
S0s between the bulge and disc for clues to the transformation process
that created that galaxy. One way of looking at the stellar
populations is to use multi-waveband photometry to measure colour
gradients over the galaxy, which provides information on the stellar
populations as younger and lower-metallicity stars tend to emit bluer
light.  \citet{Terndrup_1994}, \citet{Peletier_1996} and
\citet{Mollenhoff_2004} all found evidence of negative colour
gradients within the bulges of S0s and spirals, suggesting that redder
light, and therefore older or more metal-rich stellar populations, are
more centrally concentrated within these galaxies. Similarly,
\citet{Bothun_1990}, \citet{Peletier_1996} and \citet{Hudson_2010}
found that the discs of S0s and spirals are bluer than the bulges,
suggesting that disc galaxies either experienced more recent star
formation at larger radii \citep{deJong_1996_2}, or have higher
metallicities in their nuclear regions
\citep{Beckman_1996,Pompei_1997}.

Spectroscopic studies of galaxies have also been used to study stellar
population gradients across the bulges and discs of S0s. For example,
negative metallicity gradients and positive age gradients have been
detected in S0s by \citet{Fisher_1996}, \citet{Bell_2000},
\citet{Prochaska_2011} and \citet{Bedregal_2011}, which indicate that
the central regions of S0s contain younger and more metal rich
stars. Further evidence of recent star formation in bulge regions of
S0s has been detected by \citet{Poggianti_2001},
\citet{Ferrarese_2006}, \citet{SilChenko_2006b} and
\citet{Kuntschner_2006}, and a study by \citet{Pracy_2013} found
evidence of strong positive age gradients within the central
$\sim$~1~kpc of the bulges of `k+a' galaxies, which are thought to be
a transitional phase between spirals and S0s. Another recent study of
`k+a' galaxies by \citet{Rodriguez_2014} also found evidence that the
most recent star formation activity in these galaxies was centrally
concentrated within the disc, and that the transformation from spirals
most likely arose through gentler processes such as ram-pressure
stripping or galaxy-galaxy interactions.

Such studies of the star formation histories of S0 bulges and discs
have revealed age and metallicity gradients across the galaxies, but
fail to provide information on whether it represents a gradient within
the individual components, or whether it arises simply from the
superposition of varying amounts of bulge and disc light, where each
component contains stellar populations of distinct different ages and
metallicities. To overcome these limitations, we have developed a new
method for spectroscopic bulge--disc decomposition
\citep{Johnston_2012}, in which a high-quality spectrum of a galaxy is
cleanly separated into bulge and disc components
wavelength-by-wavelength to create two, one-dimensional spectra
representing purely the bulge and disc light. These clean spectra can
then be analyzed to determine the ages and metallicities of the bulge
and disc with minimal contamination in order to determine the sequence
of star-formation events that led to the formation of the S0.

In this paper, we set out to analyze the bulge and disc star formation 
histories spectroscopically for a sample of S0s from the Virgo Cluster, 
in order to determine the process that triggered their transformation 
from spirals. The Virgo Cluster was selected as the closest single system 
with sufficient members to undertake a systematic study of this 
transformation process. Section~\ref{sec:Observations and Data Reduction} 
describes the data set and reduction, and 
Section~\ref{sec:Spectroscopic decomposition} summarizes the method. 
The results for the stellar populations analysis,  the star formation 
timescales and chemical enrichment are discussed in 
Section~\ref{sec:Stellar Population Analysis}. The implications of these 
results for the likely evolutionary tracks followed by S0s are discussed 
in Section~\ref{sec:Summary}.

\section{Observations and Data Reduction}\label{sec:Observations and Data Reduction}

In order to study the transformation of spirals to S0s in clusters, we 
adopted a sample of 21 luminosity-selected S0s from the ACS Virgo 
Cluster Survey \citep{Cote_2004}. Details of the full sample are given 
in Table~\ref{galaxy_info}. The data set covers absolute B-band magnitudes 
in the range of $-22.3<M_{B}<-17.3$, with galaxy inclinations of greater  
than $40^{\circ}$, where $90^{\circ}$ is edge on, in order to minimize 
contamination from elliptical galaxies. 

\renewcommand{\tabcolsep}{0.55cm}
\begin{table*}
\caption{Sample of S0s from the Virgo Cluster.\label{galaxy_info}}
\begin{tabular}{ l l l l l l l}
\hline 
 Name & RA  & dec & B$_{T}$ & Exp. Time & Date  & Telescope  \\
 & ($^{h\ m\ s}$) & ($^{\circ}$\ \textasciiacute\ \textacutedbl) & & (s)\\
 (1)  &    (2)   & (3) & (4) & (5) & (6)  & (7)\\
\hline

VCC 798  & 12 25 24.04 & +18 11 25.90 & 10.09 & 4 $\times$ 900 & 2009 February 26 & Gemini-North \\
VCC 1535 & 12 34 03.10 & +07 41 59.00  & 10.61 & 2 $\times$ 900  & 2009 February 26 & Gemini-North \\
 &  &  &  & 2 $\times$ 900 & 2009 February 28 & Gemini-North \\
VCC 2095 & 12 52 56.00 & +11 13 53.00 & 11.18 & 4 $\times$ 900 & 2010 June 01 & Gemini-North \\
VCC 1062 & 12 28 03.90 & +09 48 14.00 & 11.40 & 4 $\times$ 900 & 2010 January 18 & Gemini-South \\
VCC 2092 & 12 52 17.50 & +11 18 50.00 & 11.51 & 1 $\times$ 900 & 2009 February 21 & Gemini-North \\
 &  &  &  & 3 $\times$ 900 & 2009 February 26 & Gemini-North \\
VCC 759  & 12 24 55.50 & +11 42 15.00 & 11.80 & 5 $\times$ 900 & 2010 June 02 & Gemini-North \\
 &  &  &  & 1 $\times$ 900 & 2010 July 08 & Gemini-North \\
VCC 1692 & 12 36 53.40 & +07 14 47.00 & 11.82 & 4 $\times$ 1000 & 2010 February 13 & Gemini-North \\
VCC 2000 & 12 44 31.95 & +11 11 25.10 & 11.94 & 4 $\times$ 900 & 2010 February 11 & Gemini-South \\
VCC 685  & 12 23 57.90 & +16 41 37.00 & 11.99 & 4 $\times$ 900 & 2010 March 11 & Gemini-North \\
VCC 1664 & 12 36 26.86 & +11 26 20.60 & 12.02 & 2 $\times$ 900 & 2009 April 27 & Gemini-North \\
 &  &  &  & 2 $\times$ 900 & 2009 April 28 & Gemini-North \\
VCC 944  & 12 26 50.53 & +09 35 02.00 & 12.08 & 4 $\times$ 900 & 2009 April 23 & Gemini-North \\
VCC 1938 & 12 42 47.40 & +11 26 33.00 & 12.11 & 1 $\times$ 900 & 2009 April 29 & Gemini-North \\
 &  &  &  & 3 $\times$ 900 & 2009 June 20 & Gemini-North \\
VCC 1720 & 12 37 30.61 & +09 33 18.80 & 12.29 & 3 $\times$ 1500 & 2011 May 30 & Gemini-North \\
 &  &  &  & 2 $\times$ 1500 & 2011 June 04 & Gemini-North \\
 &  &  &  & 1 $\times$ 1500 & 2011 June 20 & Gemini-North \\
VCC 1619 & 12 35 30.61 & +12 13 15.40 & 12.50 & 2 $\times$ 900 & 2009 June 20 & Gemini-North \\
 &  &  &  & 2 $\times$ 900 & 2010 February 13 & Gemini-South \\
VCC 1883 & 12 40 32.70 & +07 18 53.00 & 12.57 & 2 $\times$ 1500 & 2011 May 25 & Gemini-North \\
 &  &  &  & 1 $\times$ 1500 & 2011 May 27 & Gemini-North \\
 &  &  &  & 3 $\times$ 1500 & 2011 May 28 & Gemini-North \\
VCC 1242 & 12 29 53.49 & +14 04 07.00 & 12.60 & 6 $\times$ 900 & 2010 February 23 & Gemini-South \\
VCC 1250 & 12 29 59.10 & +12 20 55.00 & 12.91 & 1 $\times$ 900 & 2009 April 19 & Gemini-North \\
 &  &  &  & 3 $\times$ 900 & 2009 June 20 & Gemini-North \\
VCC 1303 & 12 30 40.64 & +09 00 55.90 & 13.10 & 3 $\times$ 1500 & 2010 July 13 & Gemini-North \\
 &  &  &  & 1 $\times$ 1500 & 2010 July 15 & Gemini-North \\
VCC 1913 & 12 42 10.70 & +07 40 37.00 & 13.22 & 4 $\times$ 900 & 2010 February 22 & Gemini-South \\
VCC 698  & 12 24 05.00 & +11 13 06.00 & 13.60 & 4 $\times$ 1800 & 2009 February 28 & Gemini-North \\
 &  &  &  &  1 $\times$ 1800 & 2009 April 23 & Gemini-North \\
 &  &  &  &  5 $\times$ 1800 & 2009 April 28 & Gemini-North \\
VCC 1833 & 12 40 19.65 & +15 56 07.20 & 14.54 & 2 $\times$ 900 & 2010 February 13 & Gemini-South \\

\hline
\multicolumn{7}{p{6.5in}}{
Note. Column (1): Galaxy name from \citet{Binggeli_1985}; Column (2): RA; Column (3): Declination; Column (4):
Total apparent blue-band magnitude from \citet{Binggeli_1985}; Column (5): Exposure time in seconds; 
Column (6): Date of observations; Column (7): Telescope
}
\end{tabular}
\end{table*}

The galaxies were observed along their major axis using the GMOS instruments 
\citep{Hook_2004} in long-slit mode on Gemini-North and Gemini-South between 
2008 April 24 and 2011 June 20. In a number of cases, the centres of the 
larger galaxies were offset from the middle of the slit in order to maximize 
the spatial coverage. The B1200 grating was used in combination with a  
0.5~arcsec slit. A central wavelength of $\sim{4730}$~\AA, slightly offset 
between the two sets of exposures to fill in the gaps between the chips, gave 
a wavelength range of $\sim{4300-5450}$~\AA \ with a dispersion of 
0.235~\AA~pixel$^{-1}$.  The spectral resolution was measured from the FWHM 
of the arc lines to be $\sim{1.13}$~\AA, which corresponds to a velocity 
resolution of $72\,{\rm km}\,{\rm s}^{-1}$ FWHM. Spatially, the CCDs were 
binned by 4 to give a final scale of 0.29 arcseconds pixel$^{-1}$.

A series of spectrophotometric and template stars were also observed with
the same instrumental set up, of which the details are given in
Table~\ref{star_info}. The template stars were selected to cover a range 
of spectral types in order to match the composite spectral type of the 
galaxy during the kinematic analysis.

Calibration flat fields and CuAr arc spectra were taken following each set 
of observations with the same instrumental set-up, and the spectra were 
reduced using the GMOS spectral reduction packages in 
\textsc{iraf}.\footnote{ \textsc{iraf} is distributed by the National
 Optical Astronomy Observatories, which are operated by the
 Association of Universities for Research in Astronomy, Inc., under
 cooperative agreement with the National Science Foundation} 
All the science and calibration frames were reduced by applying bias
subtraction, flat field correction, cosmic ray removal and an initial
wavelength calibration, and the three sections of each spectrum from
each CCD were joined together. The arc spectra were then used to
correct for the geometric distortions caused by the instrument optics
and to refine the wavelength solution over the whole spectrum; the
residuals of the resulting wavelength fits were generally $\sim 0.2-0.3$ \AA.

\renewcommand{\tabcolsep}{0.67cm}
\begin{table}
\begin{center}
\caption{Spectrophotometric (S) and template (T) stars.\label{star_info}}
\begin{tabular}{ l l l}
\hline 
 Name & T/S  & Spectral Class \\
\hline
HD054719 & T & K2 III \\
HD070272 & T & K5 III \\
HD072324 & T & G9 III \\
HD073593 & T & G8 IV \\
HD120136 & T & F6 IV \\
HD144872 & T & K3 V \\
HD145148 & T & K0 IV \\
HD161817 & T & A2 VI \\
Feige66 & S & - \\
Hiltner600 & S & - \\
LTT1788 & S & - \\
\hline
\end{tabular}
\end{center}
\end{table}

The wavelength-calibrated spectra were then sky subtracted and 
corrected for atmospheric extinction before flux calibration using 
the spectrophotometric standard star spectra. One standard star 
was observed for each programme of observations, and so all the galaxies
observed as part of that programme were flux calibrated with the same 
stellar spectrum.
Finally, the spectra 
from each galaxy were combined, using the measured positions of 
prominent sky lines to ensure the best possible registration of 
absolute wavelength calibration.

\section{Spectroscopic Bulge--Disc Decomposition}\label{sec:Spectroscopic decomposition}
\subsection{The method}\label{sec:method}

In order to study the star formation histories of the bulge and disc, it 
was first necessary to separate the light into individual bulge and disc 
spectra by spectroscopic bulge--disc decomposition \citep{Johnston_2012}.
In brief, this method involves taking the light profile of the galaxy at 
each wavelength in the spectrum, and fitting a bulge and disc light 
profile to this distribution in the same manner as for one-dimensional 
photometric bulge--disc decomposition. The bulge was modeled as a 
S\'ersic profile, 
\begin{equation}
I(R)=I_{e}\exp\left\{-b_{n}\left[(R/R_{e})^{1/n} - 1 \right]\right\}, 
\label{deVaucouleurs} 
\end{equation} 
where $R_{e}$ is the bulge effective radius, $I_{e}$ is the bulge effective 
surface brightness, $n$ is the s\'ersic index and $b_{n}$ is a variable 
related to $n$ \citep{Sersic_1968}, while the disc was modeled as the 
exponential profile,
\begin{equation}
I(R)=I_{0}\exp(-R/R_{0}),
\label{exponential}
\end{equation} 
where $I_{0}$ is the central surface brightness of the disc and $R_{0}$ 
is the disc scale length \citep{Freeman_1970}. When decomposing spectra, 
the light profiles were fitted from a radius of 2~arcsec in order to 
minimize distortion of the light profile from the point-spread function 
in the central regions of the galaxy.
\renewcommand{\tabcolsep}{0.8cm}
\begin{table*}
\caption{Results for the kinematics and decomposition parameters.\label{Decomposition_results}}
\begin{tabular}{ l l l l l l}
\hline 
 Name &  $R_{e}$ & $R_{0}$ & $n$  & $V_{LOS}$& $\sigma_{0}$\\
      &   [arcsec] & [arcsec] &  & [km s$^{-1}$] & [km s$^{-1}$] \\
 (1)  &    (2)   & (3) & (4) & (5) & (6) \\
\hline

VCC 798   & 8.9   $\pm\ 0.3 $  & 61.5  $\pm\ 0.6 $ & 1.87  $\pm\ 0.14 $ & 688     $\pm\ 5  $ & 181   $\pm\ 7 $\\
VCC 1062  & 8.7   $\pm\ 0.2 $  & 36.5  $\pm\ 0.6 $ & 1.30  $\pm\ 0.09 $ & 452     $\pm\ 4  $ & 199   $\pm\ 7 $\\
VCC 2092  & 8.74  $\pm\ 0.15 $ & 67    $\pm\ 2 $   & 1.41  $\pm\ 0.07 $ & 1316    $\pm\ 4  $ & 194   $\pm\ 8 $\\ 
VCC 1692  & 4.5   $\pm\ 0.4 $  & 22.4  $\pm\ 0.3 $ & 1.6   $\pm\ 0.4 $  & 1727    $\pm\ 3  $ & 217   $\pm\ 6 $\\
VCC 2000  & 2.43  $\pm\ 0.05 $ & 9.7   $\pm\ 0.7 $ & 0.91  $\pm\ 0.08 $ & 1052    $\pm\ 5  $ & 267   $\pm\ 8 $\\ 
VCC 685   & 2.87  $\pm\ 0.16 $ & 16.39 $\pm\ 0.14 $& 0.8   $\pm\ 0.2 $  & 1159    $\pm\ 5  $ & 200   $\pm\ 8 $\\ 
VCC 1664  & 9     $\pm\ 3 $    & 19.9  $\pm\ 0.3 $ & 3.5   $\pm\ 1.4 $  & 1118    $\pm\ 10 $ & 207   $\pm\ 17 $\\ 
VCC 944   & 14    $\pm\ 5 $    & 31.4  $\pm\ 0.5 $ & 3.9   $\pm\ 1.1 $  & 834     $\pm\ 4  $ & 146   $\pm\ 5 $\\ 
VCC 1720  & 7.4   $\pm\ 0.5 $  & 38.0  $\pm\ 1.7 $ & 1.9   $\pm\ 0.4 $  & 2316    $\pm\ 2  $ & 119   $\pm\ 3 $\\ 
VCC 1883  & 2.9   $\pm\ 0.4 $  & 145   $\pm\ 2   $ & 2.3   $\pm\ 0.4 $  & 1767    $\pm\ 1.3 $ & 77    $\pm\ 2  $\\ 
VCC 1242  & 3.2   $\pm\ 0.2$   & 15.7  $\pm\ 0.2 $ & 0.6   $\pm\ 0.2 $  & 1530    $\pm\ 2   $ & 80    $\pm\ 3  $\\ 
VCC 1303  & 7     $\pm\ 2 $    &  27   $\pm\ 3   $ & 2.1   $\pm\ 0.6 $  & 897     $\pm\ 4   $ & 91    $\pm\ 6  $\\ 
VCC 698   & 8     $\pm\ 4 $    &  22.4 $\pm\ 1.4 $ & 2.4   $\pm\ 1.4 $  & 2079    $\pm\ 2   $ & 52    $\pm\ 2  $\\

\hline
\multicolumn{6}{p{6.5in}}{
Note. Column (1): Galaxy name; Column (2): Bulge effective radius; Column (3):  Disc scale length; (4):
Bulge S\'ersic index; Column (5): Central line-of-sight velocity; 
Column (6): Central velocity dispersion. 
}

\end{tabular}
\end{table*}

\begin{figure*}
  \includegraphics[width=1\linewidth]{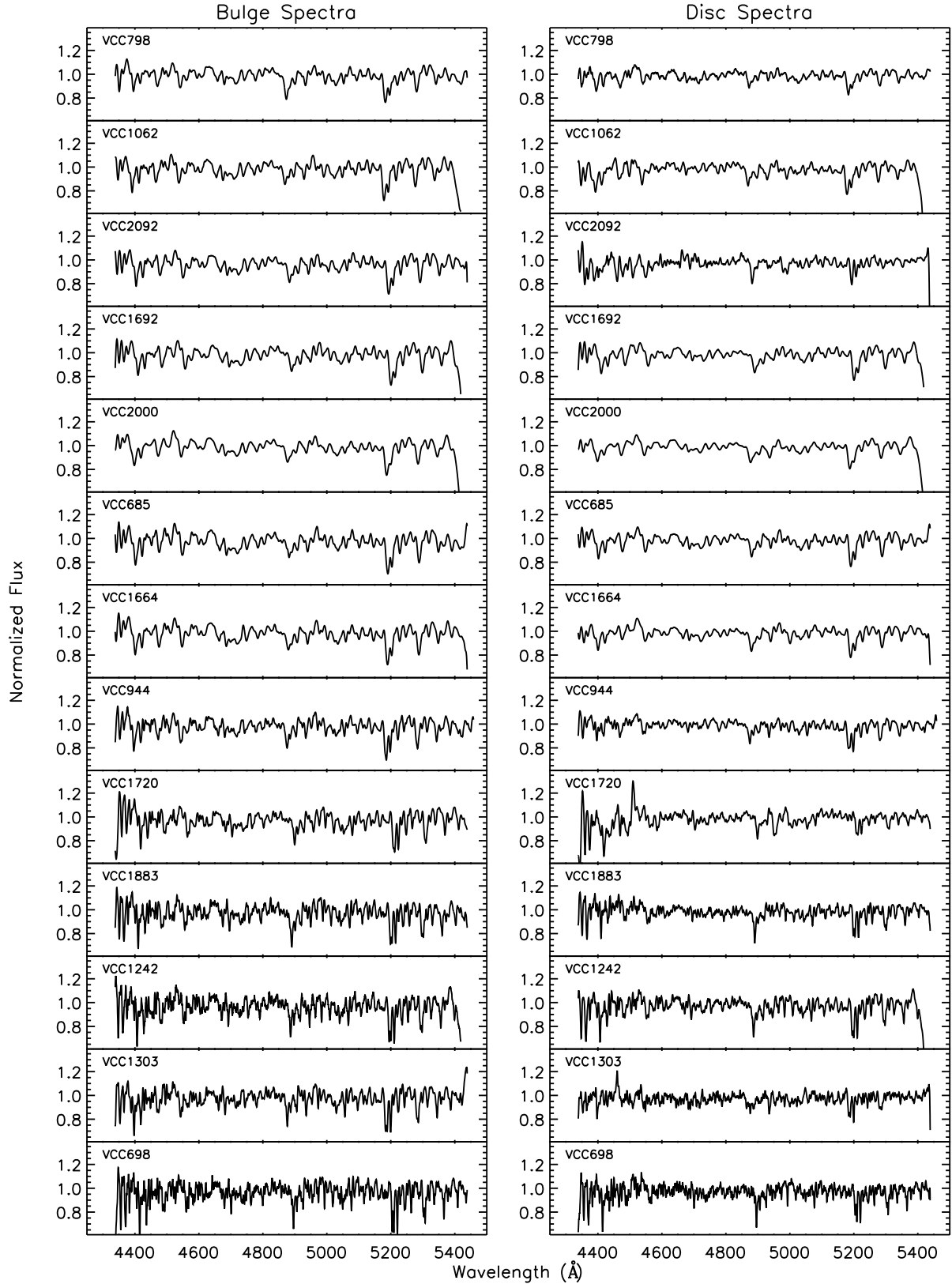}
  \caption{The decomposed bulge and disc spectra for the galaxies that could 
	    be decomposed with the Sersic bulge $+$ exponential disc model.
    \label{decomposed_spectra}}
\end{figure*}

The bulge and disc parameters at each wavelength were integrated to obtain 
the total light from each component at that wavelength, and then plotted 
against wavelength to produce the one-dimensional bulge and disc spectra.
Examples of the decomposed bulge and disc spectra for each galaxy 
in the sample are plotted in Fig.~\ref{decomposed_spectra}.
To ensure that the light profile is derived from the same rest-frame 
wavelength at all radii, thus allowing a reliable decomposition at that 
wavelength, the spectrum was first corrected for radial velocity and 
velocity dispersion variations over the radius of the galaxy. The velocity 
dispersion was equalized by convolving the spectrum at each radius with the 
appropriate Gaussian to bring it up to the maximum value measured within 
the galaxy. The radial velocity correction was then applied by measuring 
the velocity offset from the centre of the galaxy by cross correlation, and 
applying a rolling average to the shifts to produce a smooth velocity curve 
from which the correction was measured. In the outer regions of the galaxy 
where noise dominates the spectrum, the last reliable shift in velocity was 
applied. If these kinematics corrections are applied correctly, then the 
final bulge and disc spectra would be expected to have matched line-of-sight 
velocities and velocity dispersions, which would correspond to the 
line-of-sight velocity at the centre of the galaxy and the maximum velocity 
dispersion measured within the galaxy. Therefore, the kinematics of these 
spectra were tested for each galaxy using the Penalized Pixel Fitting method 
(\textsc{ppxf}) of \citet{Cappellari_2004}. \textsc{ppxf} uses the template 
stellar spectra from Table~\ref{star_info} to produce best fit models to the 
bulge and disc spectra by modeling the line-of-sight velocity distribution 
as a Gaussian with a series of Gauss-Hermite polynomials. The results for 
the bulge and disc were found to be consistent, and the values determined in 
this way are given in Table~\ref{Decomposition_results}.

\begin{figure*}
  \includegraphics[width=1\linewidth]{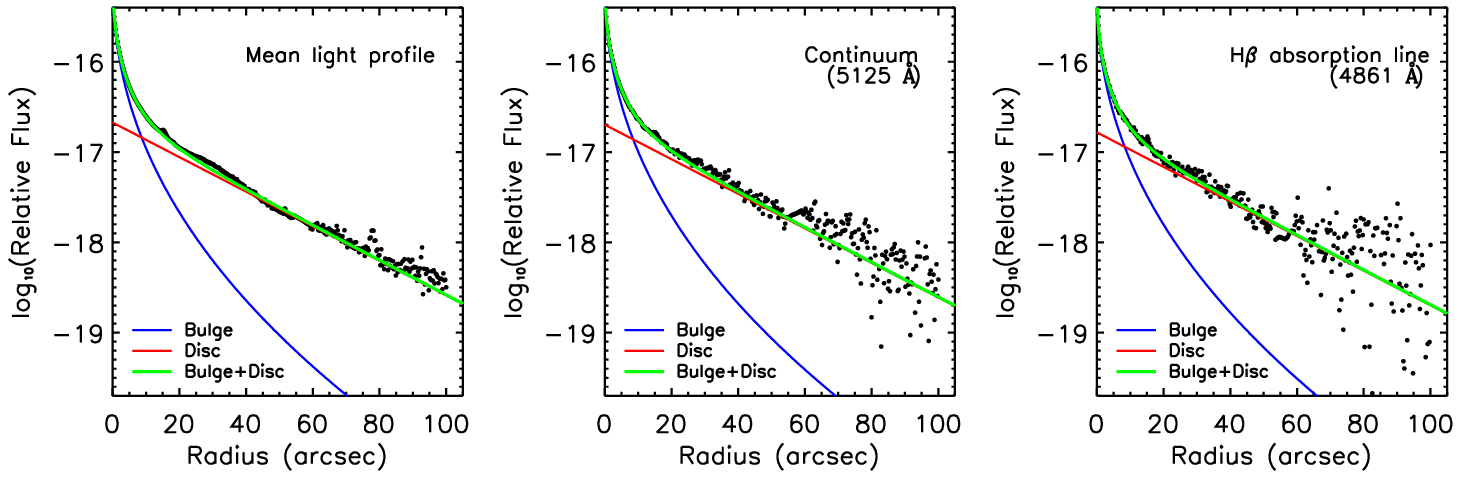}
  \caption{The best fits to the mean light profile for VCC~698 (left), 
	    and example light profiles taken from the continuum (middle) and 
	    within the H$\beta$ absorption feature (right). The bulge, 
	    disc and bulge~$+$~disc profiles are represented by the blue, 
	    red and green lines respectively.
    \label{fits}}
\end{figure*}

Many of galaxies within the sample were only slightly offset from the 
centre of the slit, allowing independent decomposition of both their 
semi-major axes. This repeated analysis was useful to check the 
reliability of the results by ensuring that they were consistent with 
each other. It was found that if all the bulge and disc parameters were 
left free for the fit, the results became very unstable, we believe due 
to a combination of degeneracy issues from leaving too many free 
parameters in the fits, and residual scattered light in the spectra. 
To reduce the degeneracy, the light 
profiles were fitted with fixed values for $R_{e}$, $R_{0}$ and 
$n$, where these quantities were measured from a fit to the mean light 
profile from the entire spectrum; the results for these
decomposition parameters are given in Table~\ref{Decomposition_results}.
Examples of the light profile fits using these parameters are 
given in Fig.~\ref{fits}, which presents examples of the best fit to the 
mean light profile for VCC~698, and the fits achieved for individual light 
profiles from the continuum and within the H$\beta$ absorption feature.
With this added constraint, the results 
from both semi-major axes became consistent with each other, and 
the co-added bulge and disc spectra bore a close resemblance to the 
original spectrum.

It was also hoped that by holding these parameters fixed in the fits, 
the effects of any residual scattered light could be mitigated. Since  
the physical distribution of the scattered light on the CCD is independent of 
the position of the galaxy on the CCD, we expected 
to obtain different results when decomposing both semi-major axes of each 
galaxy if significant amounts of scattered light was present. Therefore, 
the consistency in the results, 
both in terms of the decomposition parameters and the 
resultant stellar populations and alpha-enrichment analysis presented in 
Sections~\ref{sec:age and metallicity measurements} and \ref{sec:SF timescales}, 
suggests that the results presented here are not affected significantly by 
contamination from residual scattered light.

It was already found in \citet{Johnston_2012} that this method for
decomposition was limited to galaxies with high signal-to-noise ratios
and with clean light profiles that could be approximated by the simple
bulge + disc model. Of the 21 galaxies within the current data set,
thirteen could be decomposed reliably using this method. Of the
remaining eight, four (VCC~2095, VCC~759, VCC~1913 and VCC~1938) showed more
complicated light profiles due to the presence of dust lanes,
secondary discs, rings etc that could not be fitted by the simple
bulge + disc model used here, and the faintest galaxy in the sample,
VCC~1833, had too low a signal-to-noise ratio to fit both components
reliably. In addition, VCC~1535 and VCC~1250 also show significant
H$\beta$, [O\textsc{iii}]$_{\lambda4959}$ and
[O\textsc{iii}]$_{\lambda5007}$ emission originating from nuclear
discs in their central regions, which could not be accounted for in
the the light profiles with the simple model used in this study. The
final recalcitrant galaxy was VCC~1619, which contains two
counter-rotating stellar discs of similar mass and size, making it
unsuitable for this type of decomposition. A separate decomposition of
the stellar populations of the two discs in this bizarre galaxy was
carried out using their kinematics, which is described in
\citet{Johnston_2013}.

\subsection{Comparison with decomposition of SDSS images}\label{sec:SDSS}

The method outlined in Section~\ref{sec:method} is restricted by the
use of long-slit spectra of the major axis only, which may introduce
contamination in the stellar populations analysis due to the presence
of structures such as dust lanes, bars, rings etc. that lie in the
plane of the disc. More sophisticated methods of 2-D bulge--disc
decomposition applied to images of galaxies have the advantage that
they can work around such features and thus obtain better measurements
of the bulge and disc parameters using the full structural information
available.  It would therefore be interesting to see whether our
simpler 1-D decomposition could be compromising the extracted
spectra.  

To this end, full images of each galaxy were obtained from the SDSS
DR7 \citep{Abazajian_2009}, and mosaiced together with the
\mbox{\textsc{Montage}}
software\footnote{http://montage.ipac.caltech.edu/} to produce a large
enough field of view for a 2-D photometric decomposition. The SDSS
g-band images were selected because the central wavelength of this
band, $4770$~\AA, lies closest to the central wavelength of the
spectra, and the corrected (fpC) frames were used, and they had
already undergone bias subtraction and flat fielding as part of the
\textit{frames} pipeline \citep{Stoughton_2002}.

The decomposition was carried out using the \mbox{\textsc{Galfit}}
image analysis software \citep[v3.0.4]{Peng_2002,Peng_2010}, which is
a 2D parametric galaxy fitting algorithm. In order to compare directly
to the spectral decompositions, the SDSS images were fitted with a
S\'ersic bulge and exponential disc profile.  Each fit was also
convolved with a PSF created for each galaxy by median stacking images
of stars within the mosaiced image. The best fit model of each image
and the residuals produced by \textsc{Galfit} were checked by eye, and
the software re-run with new initial parameters if the provisional fit
was found to be poor. The results for the bulge and disc sizes and
bulge S\'ersic index were then compared with those from the spectral
decomposition, which produced a good correlation between the two
methods with a low level of scatter which could be attributable to
features within the plane of the disc as outlined above. As a further
test of the impact of such systematic distortions, the spectra were
decomposed again using fixed values for $R_{e}$, $R_{0}$ and $n$ from
the decomposition of the SDSS images and allowing only the amplitudes
of the components to vary. This test revealed little difference
between the bulge and disc ages and metallicities from the original
spectroscopic decomposition results (see Sec.~\ref{sec:Stellar
  Population Analysis}) and those decomposed using the SDSS values,
thus confirming that the spectral decomposition is fairly robust
against such modest systematic errors in the bulge and disc
parameters.

\subsection{Analysis of the systematic errors due to kinematics}\label{sec:Kinematics}

Another potential issue with the decomposition method is that the kinematic 
corrections applied before the decomposition could result in losing information
on the bulge and disc star formation histories. For example, by broadening 
the spectrum to match the maximum velocity dispersion, information on the line 
strengths in the outer regions may be lost, resulting in less reliable 
measurements of the line indices, and thus compromising the stellar populations 
analysis. Similarly, the radial velocity corrections were 
carried out by measuring the overall offset in the spectrum from the centre 
of the galaxy, which does not take into account the different rotational 
velocities that the bulge and disc will have. 

In order to test the significance of these effects when decomposing the 
corrected two-dimensional spectrum, a series of simulated spectra 
were created and decomposed in the same way as the galaxy spectra. The model 
spectra were formed by co-adding spectra of known different stellar populations 
and velocity dispersions to represent the bulge and disc, where these spectra were 
obtained from the \textsc{miles} stellar library  \citep{Sanchez_2006}. The 
proportion of bulge and disc light to add at each radius was determined by creating 
a model light profile based on typical measurements from the decomposition of the 
Virgo Cluster S0s, and the disc spectrum offset in the wavelength direction in order 
to simulate the rotational velocity of that component relative to a non-rotating 
model. As a final step, noise was added to the model spectra to simulate the 
uncertainties in the measurements. 

The simulated spectra were then decomposed in the same way as the galaxy spectra, 
and the stellar populations of the bulges and discs compared to the input values. 
The results were found to be consistent with the original stellar populations 
that went into the spectrum, thus indicating that the kinematic corrections 
described in Sec.~\ref{sec:method} are sufficient to blur the individual kinematics 
of the bulge and disc to allow successful decomposition, while minimizing 
the information lost about the strengths of any features in the data.

%

\section{Stellar Population Analysis}\label{sec:Stellar Population Analysis}
\subsection{Age and metallicity measurements}\label{sec:age and metallicity measurements}

\begin{figure}
  \includegraphics[width=1\linewidth]{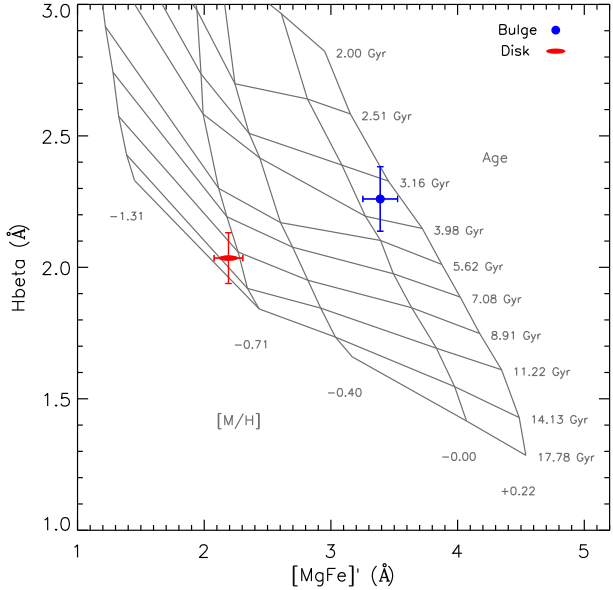}
  \caption{Example of the VCC~698 SSP model, with the line index 
	    measurements from the bulge and disc spectra over-plotted.
	   The blue circle represents the bulge while the red ellipse 
	   corresponds to the disc value. The error bars represent 
	   the statistical uncertainties.
    \label{SSP model}}
\end{figure}

\begin{figure}
  \includegraphics[width=1\linewidth]{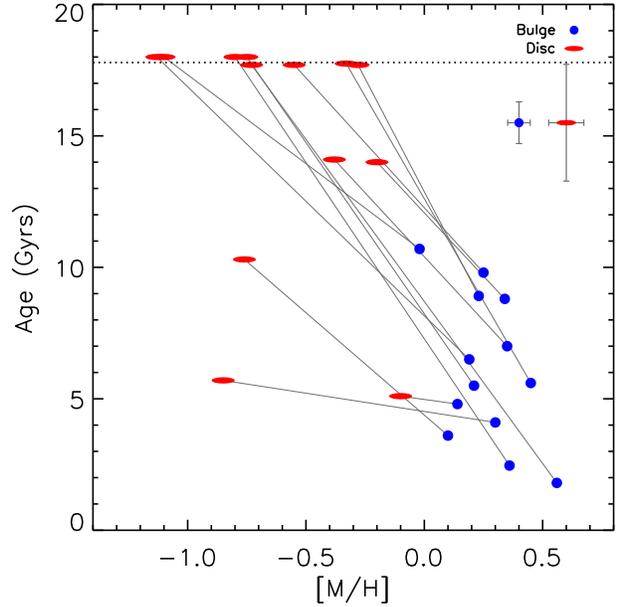}
  \caption{Estimates of the relative ages and metallicities of the bulges 
	  (blue circles) and discs (red ellipses) of the Virgo Cluster 
	  S0s. The solid lines link bulge 
	  and disc stellar populations from the 
	  same galaxy, and the horizontal dotted line links the lower edge of the 
	  SSP models used. The mean errors on the bulge and disc measurements 
	  are shown on the right of the plot.
    \label{Age-Metallicity_Virgo}}
\end{figure}

In order to extract the information on age and metallicity in the 
one-dimensional bulge and disc spectra, we used the hydrogen, magnesium 
and iron absorption line strengths as indicators of age and metallicity. 
The strengths of these absorption features were measured with the 
\mbox{\textsc{indexf}} software \citep{Cardiel_2010},
which uses the Lick/IDS index definitions to calculate a pseudo-continuum 
over each absorption feature based on the level of the spectrum in bands 
on either side, and measures the strength of the feature relative to the 
pseudo-continuum \citep{Worthey_1994, Worthey_1997}. The combined 
metallicity index, [MgFe]$'$, was then determined from these values, 
having been selected due to its negligible dependence on the 
$\alpha$-element abundance \citep{Gonzalez_1993,Thomas_2003}. The H$\beta$ 
feature was also corrected for contamination from emission by using the 
[O\textsc{iii}]$_{\lambda5007}$ emission line strength in the relation 
\begin{equation} 
\Delta(\text{H}\beta)=0.6 \times
  \text{EW}[\text{O}\text{\textsc{iii}}]_{\lambda5007}
\end{equation}
\citep{Trager_2000}, where the [O\textsc{iii}]$_{\lambda5007}$ index 
was measured from the residual spectrum obtained by subtracting the 
best combinations of stellar templates produced by \textsc{ppxf} from 
the original bulge and disc spectra. For this sample, the majority of 
the corrections were only of the order of $\sim5$~per~cent of the H$\beta$ 
index. The uncertainties in the line index measurements were estimated 
from the propagation of random errors and the effect of uncertainties 
in the line-of-sight velocities.

Single Stellar Population (SSP) models are often used to convert line 
index values into quantitative measures of age and metallicity. The SSP 
models adopted in this study are those of \citet{Vazdekis_2010}, which 
uses the \textsc{miles} stellar library.  The library spectra have a 
resolution of $\sim58.4$~km s$^{-1}$, and are convolved with a Gaussian 
of the appropriate dispersion to reproduce the spectral resolution of 
the data using a web-based tool\footnote{http://miles.iac.es/}, thus 
providing SSP models that are matched to the data and minimizing the 
loss of information that normally occurs when degrading the data to 
match lower-resolution models. In the case of VCC~698, the velocity 
dispersion of the galaxy was lower than than that of the library 
spectra, and so the decomposed spectra of this galaxy were instead 
broadened to $58.4$~km s$^{-1}$. Figure~\ref{SSP model} shows the 
example of an SSP model grid for this galaxy, with the line index 
measurements for the bulge and disc over-plotted. The global, 
luminosity-weighted ages and metallicities of the bulge and disc can be 
calculated by interpolating across the SSP model grid. Clearly in this 
example the bulge appears to contain a younger and more metal-rich 
stellar population than the disc.

This analysis was applied to all galaxies that were decomposed with the 
simple bulge~plus~disc model outlined in 
Section~\ref{sec:Spectroscopic decomposition},  with each semi-major axis 
compared independently where possible as a test of the reproducibility of 
the results. In general it was found that the line index measurements for 
the bulge and disc were consistent when compared for both halves of each 
galaxy. Hence, single measurement for the properties of the bulge and disc 
stellar populations was derived for each such galaxy by taking the mean 
value of the line indices from each semi-major axis. 
Figure~\ref{Age-Metallicity_Virgo} shows the results of this stellar 
population analysis for the decomposed Virgo Cluster galaxies, where the 
bulge results are represented by circles, the disc results by ellipses, 
and the lines link bulge and disc results from the same galaxy. Where the 
line index measurements lay off the SSP grid, the corresponding stellar 
populations were estimated by extrapolation, except where the H$\beta$ 
line index fell below the SSP model, in which case that component was 
assigned a nominal age of 18 billion years. It is important to note that 
currently different SSP models give different absolute results due to the 
remaining uncertainties in stellar astrophysics, and therefore the results 
shown in Fig.~\ref{Age-Metallicity_Virgo} should be considered as 
constraints on the relative ages and metallicities of the different stellar 
populations rather than their absolute values. The mean errors in the bulge 
and disc measurements shown in Fig.~\ref{Age-Metallicity_Virgo} are derived 
from  a combination of the difference between the stellar populations from 
both semi-major axes, the statistical uncertainties seen in 
Fig.~\ref{SSP model}, and interpolation errors when interpreting the SSP 
models.

Figure~\ref{Age-Metallicity_Virgo} clearly shows that the bulges
contain systematically younger and more metal rich stellar populations
than the discs, implying that they hosted more recent formation
activity than the discs.  Such recent central star formation
activity after the disc was quenched would explain why S0s have been
found to host positive age and negative metallicity gradients
throughout their entire structure, while the precursor spirals show
old bulges surrounded by a young disc.

\subsection{Measuring $\alpha$-enhancement}\label{sec:SF timescales}

\begin{figure}
  \includegraphics[width=1\linewidth]{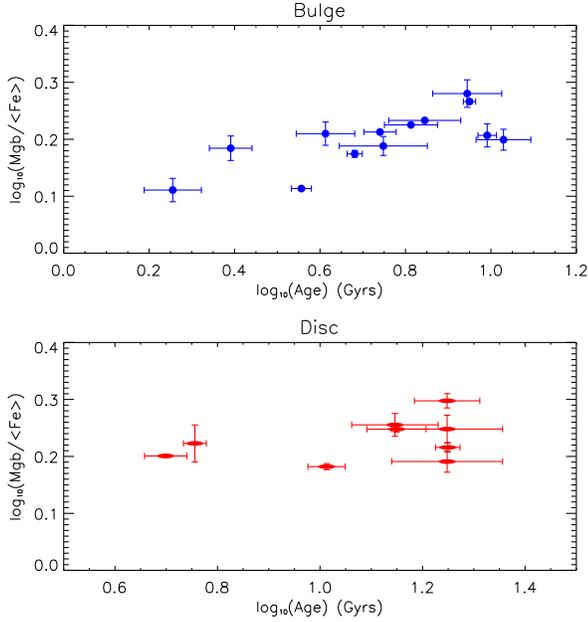}
  \caption{The bulge (top) and disc (bottom) Mgb/$\langle$Fe$\rangle$ ratios plotted 
	    against their ages. Note that in the bottom plot, galaxies with disc ages 
	    greater than the upper limit of the SSP models have been omitted.
    \label{elemental_abundances2}}
\end{figure}

A further constraint on the stellar populations of the bulges and discs is 
offered by their $\alpha$-enhancement, as this quantity provides information 
about the timescales for star formation in these two components. 
Nucleosynthesis models predict that a significant proportion of 
$\alpha$-elements present in the ISM are ejected from Type~II supernovae 
(SNe), whereas Type~Ia SNe tend to enrich the ISM with Fe \citep{Thomas_2003}. 
Since Type~II SNe start to occur shortly after star formation has begun while 
Type~Ia SNe need longer for their progenitor stars to evolve, the $\alpha$-element 
abundance tells us about the star formation timescale of a stellar 
population. 

The ratio of the Lick Mgb index over the mean of the Lick Fe5270 and
Fe5335 indices was selected as a robust proxy of the $\alpha$-element
abundance.  Short star-formation events are characterized by larger
values of this quantity, Mgb/$\langle$Fe$\rangle$, with the ratio
decreasing for star formation timescales of longer than $\sim1$~Gyr
due to the increasing Fe enrichment on such timescales.

Since this analysis directly compares measurements of the 
line indices for all galaxies simultaneously, all the decomposed bulge 
and disc spectra were broadened to match the resolution of VCC~2000, 
the galaxy with the highest velocity dispersion in this sample. Note 
that this step is not necessary when studying the stellar populations 
of each galaxy independently, as in 
Section~\ref{sec:age and metallicity measurements}. 
Figure~\ref{elemental_abundances2} 
presents the Mgb/$\langle$Fe$\rangle$ ratio for each decomposed spectrum 
plotted against the age of the bulge and disc. It can be seen that the
 bulge spectra show increasing Mgb/$\langle$Fe$\rangle$ ratios with
increasing ages; a Spearman rank test on these data shows a
correlation coefficient of 0.64, implying a greater than 98~percent
confidence that these quantities are correlated. This enhanced Fe
enrichment in the younger bulges suggests that the enriched gas that
fed the most recent star-formation events in these bulges had been
contaminated by exposure to a longer period of star formation than in
the older bulges. The disc spectra, on the other hand, show no obvious
correlations between their Mgb/$\langle$Fe$\rangle$ ratios and their
ages; since many of the discs were found to be old (see
Fig.~\ref{Age-Metallicity_Virgo}), their light will not be dominated
by the latest star-formation event, but instead represents the sum of
all the stellar populations within the entire disc, so any similar
correlation might be expected to be completely washed out.

\begin{figure}
  \includegraphics[width=1\linewidth]{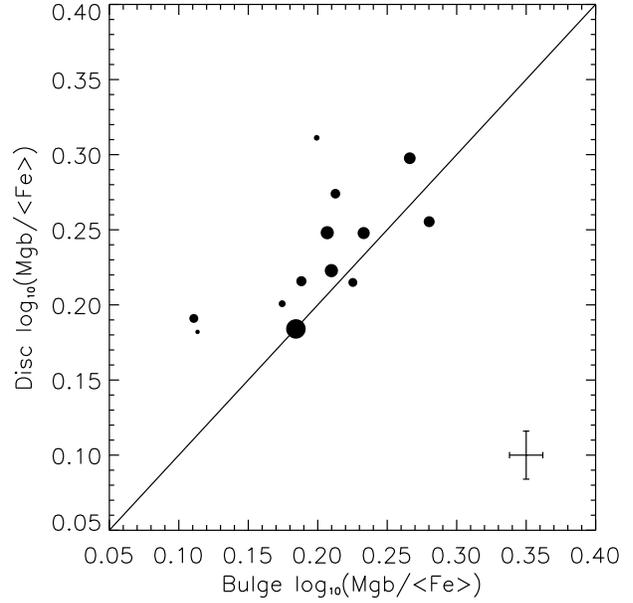}
  \caption{The Mgb$/\langle$Fe$\rangle$ ratios for the bulge 
	   and disc of each galaxy. The size of the symbol  
	   represents the luminosity of the galaxy in the K-band,  
	   where larger symbols indicate brighter galaxies. The   
	   mean error for the data points is given in the bottom right.
    \label{elemental_abundances}}
\end{figure}

\begin{figure}
  \includegraphics[width=1\linewidth]{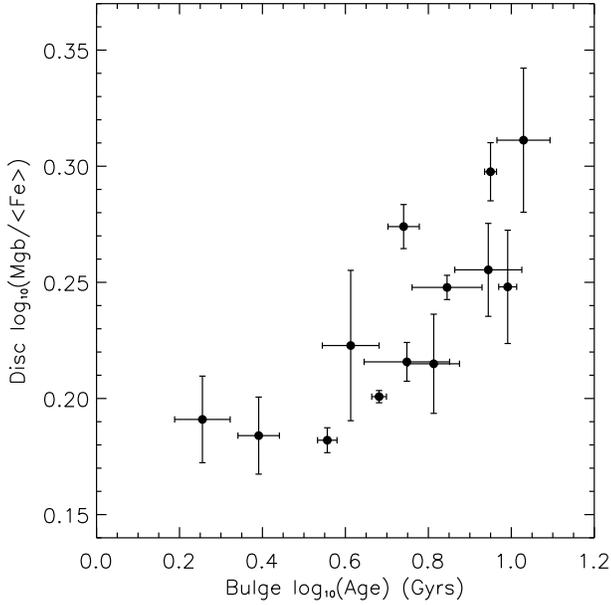}
  \caption{The relationship between the disc Mgb$/\langle$Fe$\rangle$ relative 
	  abundances and the age of the corresponding bulges, 
	  which acts as an indicator of when the final episode 
	  of star formation occurred in that galaxy.
    \label{BulgeAge_DiscMgFe}}
\end{figure}

Interestingly, a comparison of the bulge and disc
Mgb/$\langle$Fe$\rangle$ ratios, shown in
Fig.~\ref{elemental_abundances}, does reveal a correlation, 
with a correlation coefficient of 0.69 and a greater than 
99~percent significance. This correlation suggests
that the bulge and disc star formation histories are connected, but
also shows that bulges are in general more Fe enriched than the discs
of the same galaxy. This result is consistent with a scenario where
the gas that produced the final star-formation event in the bulge was
pre-enriched by earlier star formation within the disc. Further
evidence for this scenario appears in Fig.~\ref{BulgeAge_DiscMgFe},
which clearly links the age of the bulges to the
Mgb/$\langle$Fe$\rangle$ ratios of their surrounding discs, such that
galaxies with older bulges had the star formation in their discs
truncated longer ago and after a shorter timescale. 


\section{Discussion}\label{sec:Summary}
\begin{figure}
  \includegraphics[width=1\linewidth]{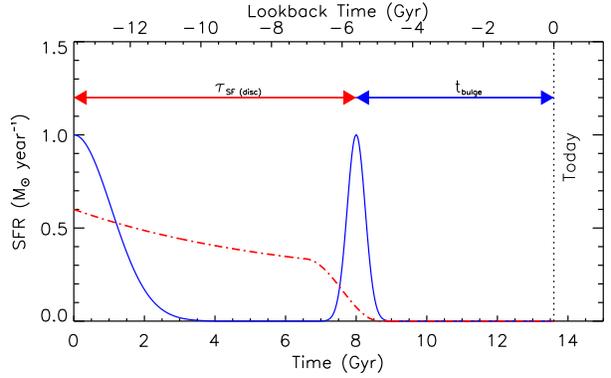}
  \caption{A simplified star formation history for the bulge and disc of 
 	  an S0 galaxy, showing the relationship between the age of the 
	  bulge stellar populations, t$_{bulge}$, and the star formation 
	  timescale of the disc, $\tau_{SF (disc)}$.
	  The disc (dot-dash line) experiences continuous star formation, 
	  the rate of which declines gradually with time,
	  until the quenching process begins, which finishes soon after 
	  when a central star-formation event uses up the remaining disc 
	  gas to produce the predominantly-young bulge of the final S0 
	  (solid line).
          \label{SF_history}}
\end{figure}

In this paper, by decomposing Virgo Cluster S0 galaxies into clean
disc and bulge spectra, we have been able to uncover a number of new
facts about these individual components as well as the connections 
between them.  From these data, a coherent quantitative picture of
each S0's star-formation history is beginning to emerge, which we
summarize in cartoon form in Fig.~\ref{SF_history}.  The galaxy starts
out as a normal spiral, with an old bulge surrounded by young,
star-forming disc. At some traumatic point in the galaxy's life, the
gas in the disc is stripped, thus quenching the star formation there,
and in the process some of the gas gets dumped in the centre of the
galaxy leading to a final burst of star formation in the bulge.  The
galaxy then fades to the S0 that we see today with a predominantly
younger and more metal rich bulge surrounded by an older and more
metal poor disc, as so clearly found in
Fig.~\ref{Age-Metallicity_Virgo}.  Although strong indications of this
phenomenon have been found previously through radial variations in age
and metallicity in S0 galaxies, this study confirms that the
phenomenon can be traced to the superposition of distinct bulge and
disc components rather than more general gradients within those
components.

A subtler probe of the star formation histories of bulges and
discs is provided by their $\alpha$-element abundances.  As we saw in
Fig.~\ref{elemental_abundances2}, there is a significant correlation
between Mgb/$\langle$Fe$\rangle$ and age in the bulges of these
galaxies, but not in their discs, which can both be understood in the
context of Fig.~\ref{SF_history}.  The emission from the bulge is
dominated by the younger stars from the final burst of star formation,
so the value of Mgb/$\langle$Fe$\rangle$ is largely dictated by the
gas from which this burst formed, which in this picture originated in
the disc and was dumped into the bulge when the galaxy transformed.
Thus, it reflects the properties of the gas in the disc at the end of
its star-forming life.  In general, the longer ago this transformation
occurred (and hence the older the age inferred for the bulge), the
shorter the star-forming lifetime of the disc because $\tau_{SF
  (disc)} + t_{\rm bulge}$ in Fig.~\ref{SF_history} reflect the total
age of the galaxy.  If $\tau_{SF (disc)}$ is relatively short (so
$t_{\rm bulge}$ is relatively long), the gas left at the end of the
disc's star-forming lifetime will not be so polluted by Fe from type~Ia 
SNe, so Mgb/$\langle$Fe$\rangle$ will be relatively large,
explaining the correlation seen.  In the disc, on the other hand, the
observed value of Mgb/$\langle$Fe$\rangle$ reflects the more extended
and potentially complicated complete star-formation history of this
component, as its light will not be dominated by a single
star-formation event, and the derived luminosity-weighted age will be
similarly complex, so the absence of any correlation in this component
is not a surprise.

This connection between the polluted gas from the disc and the visible
last burst of star formation in the bulge is underlined by
Fig,~\ref{elemental_abundances}, which shows the general trend that
the Mgb/$\langle$Fe$\rangle$ in the two components are correlated, but
that the disc is less Fe-enriched than the bulge.  This difference
arises because the disc's value for Mgb/$\langle$Fe$\rangle$ reflects
its entire star-formation history, some of which will have occurred at
early times before the Type~Ia SNe started producing large quantities
of Fe, whereas the bulge population is dominated by stars produced
from the most polluted disc gas, which will be significantly more Fe
enriched.  There is also an interesting hint in this figure that the
most massive galaxies seem to show the least difference between
Mgb/$\langle$Fe$\rangle$ for discs and bulges, which would suggest an
earlier transformation leading to less difference in the degree of
Fe enrichment, as perhaps a new example of the ``downsizing''
phenomenon.  

As a final illustration of the physics that underlies
Fig.~\ref{SF_history}, Fig.~\ref{BulgeAge_DiscMgFe} shows the clear
correlation between Mgb/$\langle$Fe$\rangle$ for the disc component
and the age of the bulge.  Again, this fits with the finite time
available for galaxy evolution, such that if the transformation occurs
later then the disc will have had time to become strongly polluted by
Fe, reducing Mgb/$\langle$Fe$\rangle$, and the bulge will have
undergone its final burst of star formation relatively recently,
decreasing its luminosity-weighted measured age.  

As this discussion indicates, there is now a wealth of information
that can be gleaned by decomposing spectra of S0 galaxies into their
bulge and disc contributions, in studying the detailed stellar
population properties of these individual components.  We are at
the point of being able not only to put together the general picture
of the quenching of disc star formation accompanied by a final episode
of bulge star formation shown in Fig.~\ref{SF_history}, but also
looking at the variations from galaxy to galaxy to tie down the
different histories that different galaxies have witnessed.  Clearly,
such decompositions would be more robust if carried out in two
dimensions using integral field unit (IFU) data rather than long-slit
spectroscopy, and the up-coming very large IFU Mapping Nearby Galaxies 
at APO (MaNGA) survey in SDSS-IV
promises the size of sample that will answer the remaining questions
about the relative importance of different transformation mechanisms.

\section*{Acknowledgements}

    We would like to thank Martin Bureau and Sugata Kaviraj for useful 
    discussions that helped us understand these results better, and Steven 
    Bamford for his help with the SDSS comparisons. We would also 
    like to thank the anonymous referee for their useful comments that helped 
    improve this paper. This work was based on observations obtained  
    at the Gemini Observatory, which is operated by the Association of Universities
    for Research in Astronomy, Inc., under a cooperative agreement with the
    NSF on behalf of the Gemini partnership: the National Science
    Foundation (United States), the Science and Technology Facilities
    Council (United Kingdom), the National Research Council (Canada),
    CONICYT (Chile), the Australian Research Council (Australia),
    Minist\'{e}rio da Ci\^{e}ncia, Tecnologia e Inova\c{c}\~{a}o (Brazil)
    and Ministerio de Ciencia, Tecnolog\'{i}a e Innovaci\'{o}n Productiva
    (Argentina). The programme IDs were GN-2008-Q-105, GN-2009A-Q-102,
    GN-2010A-Q-60 and GS-2010A-Q-23.

   This research made use of Montage, funded by the National Aeronautics 
   and Space Administration's Earth Science Technology Office, Computation 
   Technologies Project, under Cooperative Agreement Number NCC5-626 between 
   NASA and the California Institute of Technology. Montage is maintained 
   by the NASA/IPAC Infrared Science Archive


Funding for the SDSS and SDSS-II has been provided by the Alfred P. 
Sloan Foundation, the Participating Institutions, the National Science Foundation, 
the U.S. Department of Energy, the National Aeronautics and Space Administration, 
the Japanese Monbukagakusho, the Max Planck Society, and the Higher Education 
Funding Council for England. The SDSS Web Site is http://www.sdss.org/.

The SDSS is managed by the Astrophysical Research Consortium for the Participating 
Institutions. The Participating Institutions are the American Museum of Natural 
History, Astrophysical Institute Potsdam, University of Basel, University of 
Cambridge, Case Western Reserve University, University of Chicago, Drexel University, 
Fermilab, the Institute for Advanced Study, the Japan Participation Group, Johns 
Hopkins University, the Joint Institute for Nuclear Astrophysics, the Kavli Institute 
for Particle Astrophysics and Cosmology, the Korean Scientist Group, the Chinese 
Academy of Sciences (LAMOST), Los Alamos National Laboratory, the Max-Planck-Institute 
for Astronomy (MPIA), the Max-Planck-Institute for Astrophysics (MPA), New Mexico State 
University, Ohio State University, University of Pittsburgh, University of Portsmouth, 
Princeton University, the United States Naval Observatory, and the University of Washington.

EJ acknowledges support from the STFC and the RAS.

\footnotesize{
\bibliographystyle{mn2e}

\bibliography{paper3_refs}
}

\end{document}